\documentclass[ aps, longbibliography, twocolumn, reprint, showpacs, nofootinbib, superscriptaddress]
{revtex4-1}
\usepackage{graphicx}
\usepackage{amssymb}   
\usepackage{amsfonts}
\usepackage{amsmath}
\usepackage{dsfont}
\usepackage{dcolumn}   
\usepackage{bm}        
\usepackage[mathscr]{eucal}
\usepackage[dvipsnames]{xcolor}
\usepackage[colorlinks, linkcolor=WildStrawberry,citecolor=WildStrawberry,urlcolor=PineGreen]{hyperref}
\usepackage[all]{hypcap} 
\usepackage{mathtools}
\usepackage{wasysym}
\usepackage{tikz}
\usepackage{array}

\newcommand{\ra}{\rangle}
\newcommand{\la}{\langle}

\newcommand{\beq}{\begin{equation}}
\newcommand{\eeq}{\end{equation}}

\newcolumntype{P}[1]{>{\centering\arraybackslash}p{#1}}
\newcolumntype{M}[1]{>{\centering\arraybackslash}m{#1}}

\begin{document}

\title{Data-driven discovery of statistically relevant information in quantum simulators}

\author{R.~Verdel}
\email{rverdel@ictp.it}
\affiliation{The Abdus Salam International Centre for Theoretical Physics (ICTP), Strada Costiera 11, 34151 Trieste, Italy}

\author{V.~Vitale}
\affiliation{The Abdus Salam International Centre for Theoretical Physics (ICTP), Strada Costiera 11, 34151 Trieste, Italy}
\affiliation{SISSA---International School of Advanced Studies, Via Bonomea 265, 34136 Trieste, Italy }
\affiliation{Universit\'e Grenoble Alpes, CNRS, Laboratoire de Physique et
Mod\'elisation des Milieux Condens\'es (LPMMC), Grenoble 38000, France}

\author{R.~K.~Panda}
\affiliation{The Abdus Salam International Centre for Theoretical Physics (ICTP), Strada Costiera 11, 34151 Trieste, Italy}
\affiliation{SISSA---International School of Advanced Studies, Via Bonomea 265, 34136 Trieste, Italy }
\affiliation{INFN Sezione di Trieste, Via Valerio 2, 34127 Trieste, Italy}

\author{E.~D.~Donkor}
\affiliation{The Abdus Salam International Centre for Theoretical Physics (ICTP), Strada Costiera 11, 34151 Trieste, Italy}
\affiliation{SISSA---International School of Advanced Studies, Via Bonomea 265, 34136 Trieste, Italy }

\author{A.~Rodriguez}
\affiliation{The Abdus Salam International Centre for Theoretical Physics (ICTP), Strada Costiera 11, 34151 Trieste, Italy}
\affiliation{Dipartimento di Matematica e Geoscienze, Universitá degli Studi di Trieste, via Alfonso Valerio 12/1, 34127, Trieste, Italy }

\author{S.~Lannig}
\affiliation{Kirchhoff-Institut f{\"u}r Physik, Universit{\"a}t
Heidelberg, Im Neuenheimer Feld 227, 69120 Heidelberg, Germany}

\author{Y.~Deller}
\affiliation{Kirchhoff-Institut f{\"u}r Physik, Universit{\"a}t
Heidelberg, Im Neuenheimer Feld 227, 69120 Heidelberg, Germany}

\author{H.~Strobel}
\affiliation{Kirchhoff-Institut f{\"u}r Physik, Universit{\"a}t
Heidelberg, Im Neuenheimer Feld 227, 69120 Heidelberg, Germany}

\author{M.~K.~Oberthaler}
\affiliation{Kirchhoff-Institut f{\"u}r Physik, Universit{\"a}t
Heidelberg, Im Neuenheimer Feld 227, 69120 Heidelberg, Germany}

\author{M.~Dalmonte}
\affiliation{The Abdus Salam International Centre for Theoretical Physics (ICTP), Strada Costiera 11, 34151 Trieste, Italy}
\affiliation{SISSA---International School of Advanced Studies, Via Bonomea 265, 34136 Trieste, Italy }


\begin{abstract}

Quantum simulators offer powerful means to investigate strongly correlated quantum matter. However, interpreting measurement outcomes in such systems poses significant challenges. Here, we present a theoretical framework for  information extraction in synthetic quantum matter, illustrated for the case of a quantum quench in a spinor Bose-Einstein condensate experiment. Employing non-parametric unsupervised learning tools that provide different measures of information content, we demonstrate a theory-agnostic approach to identify dominant degrees of freedom. This enables us to rank operators according to their relevance, akin to effective field theory. 
To characterize the corresponding effective description, we then explore the intrinsic dimension of data sets as a measure of the complexity of the dynamics. 
This reveals a  simplification of the data structure, which correlates with the emergence of time-dependent universal behavior in the studied system. Our assumption-free approach can be immediately applied in a variety of experimental platforms.

\end{abstract}

\maketitle

\section{Introduction} \label{sec:intro}

Recent remarkable advances in highly controlled synthetic quantum devices  have revolutionized the study of strongly correlated systems~\cite{RevModPhys.80.885, 10.1093/acprof:oso/9780199573127.001.0001, RevModPhys.86.153, doi:10.1126/science.aal3837, Daley2022, doi:10.1021/acs.chemrev.8b00803}. 
A key element of many of such platforms is their capacity to produce large data sets of many-body snapshots, for example, via generalized projective measurements of the entire wave function~\cite{PhysRevLett.123.063603}.
However, the analysis of such outcome poses in general serious challenges, which typically force us to rely on assumptions for certain quantities, disregarding part of the information content of the generated data---in data science language, a dimensional reduction with an uncontrolled loss of information.
A particularly important problem is the identification of the most informative observables to describe such quantum many-body systems---a paramount task at the core of quantum field theory~\cite{tsvelik_2003, doi:10.1142/10545}, that is even more daunting for systems driven out of equilibrium. 
To address this, one needs to develop methods to process the maximum amount of information in quantum simulator output, which are able to identify relevant features---and thus degrees of freedom---emerging from the underlying physical system, without making any assumption nor uncontrolled dimensional reduction.

\begin{figure*}
    \centering
    \includegraphics[width=1.\linewidth]{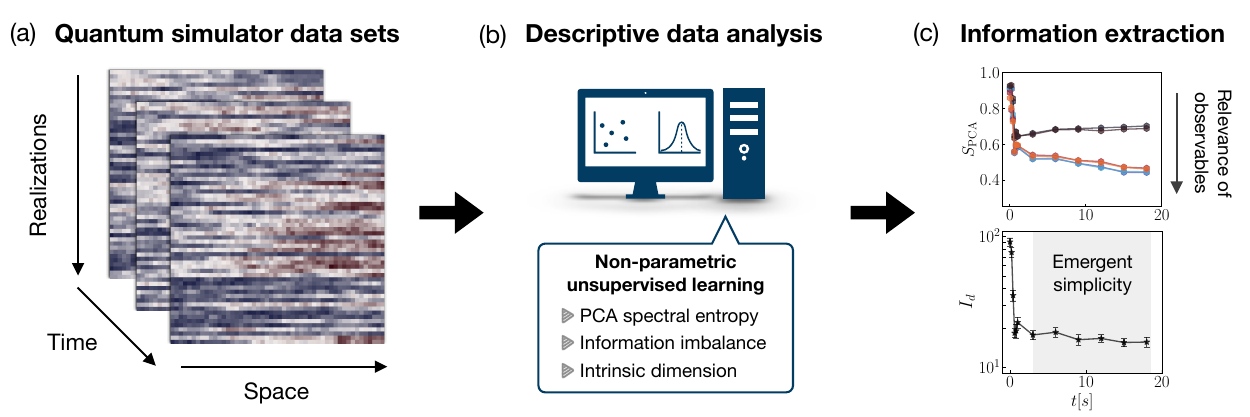}
    \caption{
    {  Assumption-free unveiling of relevant information in quantum simulation.} (a) We start from snapshots of a many-body system, which are represented as 2D arrays at different times. At a fixed time, each row corresponds to a different realization, while each column is a different data feature, e.g., the atomic density in a given magnetic substate at a given spatial location. (b) Using non-parametric unsupervised learning tools, we perform an exploratory analysis to uncover interesting features of the data, without making any assumptions.
    (c) From this description, we infer relevant properties of the physical system.  
    (Top) By quantifying the information content and correlations in a data set, the principal component analysis entropy provides a measure of relevance of observables, thence guiding the identification of the most informative degrees of freedom.
    (Bottom) After a quick fall to relatively small values, the intrinsic dimension of data sets features a long, stable plateau as a function of time (shaded region), providing a lower bound for the timescale after which the dynamics may become simpler and be captured by universal scaling.}
    \label{fig:1}
\end{figure*}

In this work, we introduce a theoretical framework for data-driven information discovery in quantum simulation, which is schematically illustrated in Fig.~\ref{fig:1}.
We start by considering collections of independent quantum simulator snapshots, which resolve, for example, the dynamics of a many-body system in space and time [Fig.~\ref{fig:1}(a)]. 
Such data sets are characterized using non-parametric unsupervised learning methods [Fig.~\ref{fig:1}(b)].
Finally, from this system-agnostic and unsupervised description of the data, we infer relevant information for the physical system under study [Fig.~\ref{fig:1}(c)].

This framework is based on three techniques: (i) spectral entropies calculated from a principal component analysis (PCA) of the data, 
(ii) the information imbalance between a subset and the full set of measured quantities, and (iii) the intrinsic dimension of the concomitant data manifolds.
These tools which quantify---from different angles---the information content and correlations in the data, have found several successful applications in various fields, such as chemical and biomolecular science~\cite{doi:10.1073/pnas.97.18.10101,10.1093/bioinformatics/btl214, 10.1093/bioinformatics/btm528, 10.1093/pnasnexus/pgac039, doi:10.1021/acs.jctc.2c01205, Facco2017, 10.1371/journal.pcbi.1006767, Allegra2020, 10.1021/acs.chemrev.0c01195}, ecology~\cite{10.3389/fevo.2021.623141}, stock market dynamics~\cite{CARAIANI2014571, GU2015103, GU2016150, Chakraborti_2021, ALVAREZRAMIREZ2021126337, ESPINOSAPAREDES2022112238}, and image analysis~\cite{NIPS2004_74934548,8953348, pope2021the, 10009680}.

To demonstrate the capabilities of our approach we apply it to experimental data of a spinor Bose-Einstein condensate (BEC) \cite{Prufer2020}: we evaluate the full set of experimentally measured densities without knowledge of the post-processing steps which are necessary in order to infer the relevant spin variables from them.
Our main results are as follows: 
(I) PCA spectral entropies and information imbalance allow for a theory-agnostic determination of the most informative measured observables.
The predictive power of these methods is demonstrated by showing that they can also unveil combinations of the measured densities, which are key to describe the spin structure of the system~\cite{Prufer2018, Prufer2020, RevModPhys.85.1191, KAWAGUCHI2012253}.
(II) The behavior of the intrinsic dimension as a function of time, displays a rapid decay to significantly smaller values, after which it features very long, stable plateaus. 
As  argued below, this observation is in strong agreement with the formation of spin structure and the emergence of self-similar dynamics~\cite{PhysRevLett.114.061601,Prufer2018, Erne2018, Prufer2020, Glidden2021}.
The remaining of this paper is structured as follows: In Sec.~\ref{sec:data_mining}, we introduce the different data science methods used in this work. We describe the spinor BEC experiment and the structure of our data sets in Sec.~\ref{sec:experiments}.  In Sec.~\ref{sec:relevance}, we present a theory-agnostic scheme to identify relevant fields, in a systematic and unbiased way, directly from (a limited number of) experimental observations. In Sec.~\ref{sec:complexity}, we complement our data-driven analysis by characterizing the Kolmogorov complexity of the studied quantum dynamics, and address its capability of recognizing physical information. Conclusions and possible extensions of our work are discussed in Sec.~\ref{sec:conclusions}.


\section{Methods} \label{sec:data_mining}

Before diving into the central part of our paper, we present an introduction to the data science tools that are employed in this work. We focus on non-parametric methods that are oriented towards extracting information from data, without making (strong) assumptions on the functional form of the probability distribution underlying the data. For the purpose of this exposition, we consider here an abstract data set, structured as a rectangular matrix ${\bf X}=\{\vec{X}^i\}_{i=1}^{N_r}$, of dimension $N_r \times p$, where each $p$-dimensional row vector $\vec{X}^i$ represents a single \emph{realization} (observation) of a set of $p$ \emph{features} (input variables) that are measured in all realizations. This type of structure is ubiquitous to both stochastic Monte Carlo simulations, as well as quantum experiments featuring projective measurements of a large number of degrees of freedom. In the next section, we shall define precisely the concrete data sets that will be analysed with the techniques presented below.

\subsection{Principal component analysis entropy} \label{subsec:S_PCA}

Principal component analysis is one of the most popular non-parametric methods for unsupervised learning, and has found a vast number of applications, including classical and quantum many-body problems~\cite{PhysRevB.94.195105, PhysRevE.95.062122, PhysRevB.96.195138, PhysRevE.96.022140, PhysRevB.96.144432, PhysRevE.97.013306, Khatami_2019, PhysRevB.106.144313}. 
The central idea of PCA is to use an orthogonal transformation to seek for directions along which the data exhibit most  variation~\cite{doi:https://doi.org/10.1002/0470013192.bsa501, MEHTA20191}. 
This is motivated by empirical evidence showing that in many cases such ``high-variance'' directions capture the relevant information of  the data.
This problem reduces to diagonalizing the sample covariance matrix ${\bf \Sigma}={{\bf X}^\star}^T {\bf X}^\star/(N_r-1)$, where ${\bf X}^\star$ is the column-centred data matrix in which the mean value of each column is subtracted from the entries in the column.
The solution to the eigenvalue-eigenvector problem ${\bf \Sigma} \vec{w}_k =\lambda_k \vec{w}_k$, yields $\lambda_1\ge \lambda_2 \ge \cdots \ge \lambda_R \ge0$ ($R \le \mathrm{min}\{N_r, p\}$ is the rank of ${\bf X}^\star$), and the normalized eigenvectors $\vec{w}_k$.  
The eigenvalues $\lambda_k$ are the variances of the principal components (PCs), which are determined by the  eigenvectors as $\vec{p}_k={\bf X^\star} \vec{w}_k$.
We note that the procedure above is equivalent to performing  a singular value decomposition (SVD) on ${\bf X}^\star$, and that the semi-definite positiveness of ${\bf \Sigma}$ then follows from the fact that the $\lambda_k$ are proportional to the squared singular values of ${\bf X}^\star$~\cite{doi:https://doi.org/10.1002/0470013192.bsa501, MEHTA20191}. 
A standard measure of importance of a given PC is given by  the corresponding normalized eigenvalue $\tilde{\lambda}_k:=\lambda_k/\sum_l \lambda_l$, that is, the  proportion of total variance that is accounted for by the $k$-th PC.

The PCA algorithm described above forms the basis of a dimensional reduction scheme, for situations in which  the first \emph{few} PCs capture most of the variation present in \emph{all} of the original variables~\cite{doi:https://doi.org/10.1002/0470013192.bsa501, MEHTA20191}. 
However, determining in a systematic way how many PCs can be disregarded without significant information loss is, in general, a difficult task.
Instead of dealing with this aspect of PCA, our goal here is to leverage \emph{all} the information contained in the PC decomposition. 
To this end, we introduce an information theoretic-inspired quantity built from the full (normalized) PCA spectrum as follows. 
Since   $\tilde{\lambda}_k\geq 0$ for all $k$, and by construction $\sum_{k=1}^R \tilde{\lambda}_k=1$,  we can regard the set of all normalized eigenvalues $\{\tilde{\lambda}_k\}$, as a probability distribution and define  the \emph{PCA entropy} in analogy to Shannon's entropy~\cite{6773024}, namely,
\beq
\label{eq:1}
S_\mathrm{PCA}(\{\tilde{\lambda}_k\}):=-\frac{1}{\ln(R)}\sum_{k=1}^R \tilde{\lambda}_k \ln (\tilde{\lambda}_k),
\eeq
where for convenience we have normalized this entropy by its maximum possible value, $\ln (R)$. 
The PCA entropy in Eq.~\eqref{eq:1}, provides a tool to explore the \emph{informational} aspect of the PC decomposition. 
Indeed, $S_\mathrm{PCA}$ measures how spread is the information on the principal axes: when no principal direction represents a preferential direction of information accumulation (i.e., $\tilde{\lambda}_k\approx 1/R$, for all $k$), then $S_\mathrm{PCA}\approx 1$, whereas, if information accumulates around some of the principal directions, then $S_\mathrm{PCA}<1$.
In fact, in the extreme case in which a single PC explains almost the full variation of the data (i.e., $\tilde{\lambda}_1\approx 1$), then $S_\mathrm{PCA}\approx0$. 
Importantly, $S_\mathrm{PCA}<1$ implies the presence of correlations among the original variables. 
We note, however, that the opposite is in general not true. 
Consider, for example, a set of two-dimensional data points randomly distributed around a circle:  none of the principal directions explains more variation of the data than the other and therefore $S_\mathrm{PCA}\approx 1$,  in spite of the input variables exhibiting a clear correlation between them.
As mentioned in the introduction, the concept of PCA entropy (and the closely related `SVD entropy') has found several applications, ranging from unsupervised feature selection methods in bioinformatics~\cite{10.1093/bioinformatics/btl214, 10.1093/bioinformatics/btm528}, to schemes to characterize complexity in ecological networks~\cite{10.3389/fevo.2021.623141} and financial signals~\cite{CARAIANI2014571, GU2015103, GU2016150, Chakraborti_2021, ALVAREZRAMIREZ2021126337, ESPINOSAPAREDES2022112238}. However, very little is known about its predictive power in the context of the many-body problem.

\subsection{Information imbalance} \label{subsec:IB}

Another recently introduced method to quantify information content goes under the name of \emph{information imbalance}~\cite{10.1093/pnasnexus/pgac039}. 
More specifically, this method quantifies the relative information retained when using two distance measures, built with different subsets of data features. 
In a physical context, this technique can therefore provide an ideal tool to systematically compare---in a fully data-driven manner---different \emph{observables} (subsets of features) and determine which of those can describe better the full space of measured quantities. 

The information imbalance method is briefly explained in the following. (The reader is referred to Ref.~\cite{10.1093/pnasnexus/pgac039} for a more detailed explanation.)
Given two distance measures  $D_A(\vec{X}^i, \vec{X}^j)$ and $D_B(\vec{X}^i, \vec{X}^j)$, defined on the same data space, we can rank  the neighbors of a point $\vec{X}^i$, by sorting, from smallest to largest, the pairwise distances between such a point and the rest of points using the two distance measures.
These rankings are encoded in the so-called rank matrices $R_{A/B}^{ij}$. Hence, $R_A^{ij}=1$, means that $\vec{X}^j$ is the 1st nearest neighbor of $\vec{X}^i$ in space $A$, and so on.
Here, we restrict ourselves to the case in which the two considered distance measures refer to the Euclidean distance computed with two subsets of features.  For example, for data points in two dimensions with components $(x^i, y^i)$, two possible choices of the distance measures are $D_A(\vec{X}^i, \vec{X}^j)=|x^i-x^j|$ and   $D_B(\vec{X}^i, \vec{X}^j)=\sqrt{(x^i-x^j)^2+(y^i-y^j)^2}$.
The key insight of the information imbalance method is the fact that the full correlation structure between the two metrics under study is essentially captured by the conditional rank distribution $p(R_B|R_A=1)$, that is, the probability distribution of the ranks $R_B^{ij}$ in space $B$ restricted to pairs of points $(\vec{X}^i,\vec{X}^j)$, such that $R_A^{ij}=1$ (i.e., nearest-neighbor points according to $A$). 
Then, the closer this distribution is to a delta function peaked at 1,  the more information about space
$B$ is contained in space $A$. 
The deviation of $p(R_B|R_A=1)$ from such a delta function is quantified by the conditional expectation $\la R_B | R_A=1\ra$~\cite{10.1093/pnasnexus/pgac039}, which is used to define the information imbalance from space $A$ to space $B$, namely
\beq
\label{eq:2}
\Delta(A \to B) = \frac{2}{N_r} \la R_B |R_A=1 \ra \approx \frac{2}{N_r^2} \sum_{i,j:R^{ij}_A=1} R^{ij}_B.
\eeq
In the limit case in which nearest neighbors in $A$ are exactly the same as those in $B$, we have that $\sum_{j:R^{ij}_A=1} R^{ij}_B= 1 $ (for a given $i$), and hence $\sum_{i,j:R^{ij}_A=1} R^{ij}_B = N_r$. Therefore, the information imbalance in Eq.~\eqref{eq:2}, vanishes as $1/N_r$. A vanishing information imbalance thus indicates that $A$ can fully predict $B$, in the sense specified above.  In the extremely opposite case in which the distance ranks estimated with the two metrics are completely uncorrelated, we  have that $\sum_{j:R^{ij}_A=1} R^{ij}_B=\frac{1}{N_r-1} \cdot \frac{1}{2}N_r(N_r-1) = \frac{N_r}{2}$  and hence $\sum_{i,j:R^{ij}_A=1} R^{ij}_B = \frac{N_r^2}{2}$. Therefore, in this case $\Delta(A \to B)=1$, and we say that `$A$ is not informative of $B$'.
A scheme for feature selection can then be carried out by measuring the information imbalance from a space of a subset of features to the space of all features. A similar approach has been used, for instance, in Ref.~\cite{doi:10.1021/acs.jctc.2c01205} to compare the information that is captured by  different atomic descriptors with respect to standard order parameters (and vice-versa) in molecular systems.

\subsection{Intrinsic dimension} \label{subsec:Id}

To complement our data analysis we consider a key concept in the sub-field of manifold learning, namely, the \emph{intrinsic dimension} ($I_d$).
The $I_d$ quantifies the least number of functionally independent parameters needed to describe the data~\cite{TRUNK1968508, Campadelli2015, CAMASTRA201626}. 
This quantity has a deep connection with information theory, for it serves as a  
proxy of the Kolmogorov or algorithmic complexity\footnote{Intuitively, the Kolmogorov complexity measures the complexity of a string as the length of the shortest computer program (in a predefined programming language) that outputs the string.} ~\cite{kolmogorov,STAIGER1993159,  vitale2023topological, 2023arXiv230113216M}.
Beyond this informational aspect, the notion of intrinsic dimension plays an important role in unsupervised machine learning, as exemplified in various applications ranging from biomolecular science~\cite{Facco2017, 10.1371/journal.pcbi.1006767, Allegra2020, 10.1021/acs.chemrev.0c01195},  to image analysis~\cite{NIPS2004_74934548,8953348, pope2021the, 10009680} .
Only recently has this concept been employed in physics, more specifically, in the study of critical behavior---in and out of equilibrium---in classical and quantum statistical mechanics systems~\cite{PhysRevX.11.011040, PRXQuantum.2.030332, PhysRevB.106.144313, 2023arXiv230113216M}.
Estimating $I_d$ is, in general, a far-from-trivial task and, in fact, an active field of research~\cite{Campadelli2015, CAMASTRA201626}.
Here we use a distance-based $I_d$ estimator that leverages information of only local neighborhoods, namely, the TWO-NN algorithm~\cite{Facco2017}, which we briefly describe in the following. (The reader is referred to Ref.~\cite{Facco2017} for a more in-depth discussion.) For each point  $\vec{X}^i$ in a generic data set, we compute the distance to its first and second nearest neighbors, denoted by $r^i_{1}, r^i_{2}$.
Next, we define the ratio $\mu^i := r^i_{2}/r^i_{1}$. 
For data that are locally uniformly distributed on a $I_d$-dimensional hypersphere, the probability distribution function of $\mu$ is given by $f(\mu)=I_d \mu^{-I_d-1}$.
The cumulative distribution function $F(\mu)$, obtained upon integration, then satisfies
\beq
\label{eq:3}
-\ln[1-F(\mu)]= I_d \ln(\mu),
\eeq
which is used to estimate $I_d$ through a linear fit of the points $\{(\ln(\mu), -\ln[1-F_\mathrm{emp}(\mu)])\}$, where $F_\mathrm{emp}(\mu)$ is the empirical cumulate.
In practice, verifying a linear relation between $\ln(\mu)$  and $-\ln[1-F_\mathrm{emp}(\mu)])$, serves also as a good consistency check of the (mild) assumption of local uniformity of the data set.
\section{Quantum simulation on a spinor BEC and experimental data sets} \label{sec:experiments}

We consider the dynamics realized by a BEC of $^{87}$Rb in the $F=1$ hyperfine spin ground state manifold confined in a quasi-one-dimensional elongated harmonic dipole trap (the data evaluated here are taken from Ref.~\cite{Prufer2020}; see this publication for further details on the experiment).
The system is initialized with all atoms in the magnetic substate $m_\text{F}=0$. 
By instantaneously changing a control parameter we tune spin-changing collision processes into resonance. This procedure implements a quench across a quantum phase transition which brings the system far from equilibrium. 

For different times $t$ after the quench we simultaneously infer the two orthogonal spin projections $F_x$ and $F_y$ from the observed densities with spatial resolution along the longitudinal trap direction \cite{PhysRevLett.123.063603} via
\begin{align}
\label{eq:4}
  F_x&=\left(n_{2,+2}-n_{2,-2}\right)/\left(n_{2,+2}+n_{2,0}+n_{2,-2}\right), \nonumber\\
  F_y&=\left(n_{1,+1}-n_{1,-1}\right)/\left(n_{1,+1}+n_{1,0}+n_{1,-1}\right),  
\end{align}
where $n_{F, m_F}$ is the density in the state with hyperfine manifold $F$ and  magnetic sublevel $m_F$. This is achieved by first performing a $\pi/2$ spin rotation around the $y$-axis to map the $F_x$-projection to the $z$-axis, which allows its detection via density differences. Then, by transferring half of the population of each $m_\text{F}$ level from $F=1$ to $F=2$, the $F_x$-projection is stored in the populations of $F=2$. Finally, another $\pi/2$ spin rotation around the $x$-axis, which exclusively addresses the $F=1$ manifold, maps the $F_y$-projection to a detectable population difference in $F=1$. All $6$ density distributions are read out with spatial resolution along the longitudinal trap direction via Stern-Gerlach separation and absorption imaging.

\begin{figure*}[bt!]
	\centering
	\includegraphics[width=1.\textwidth]{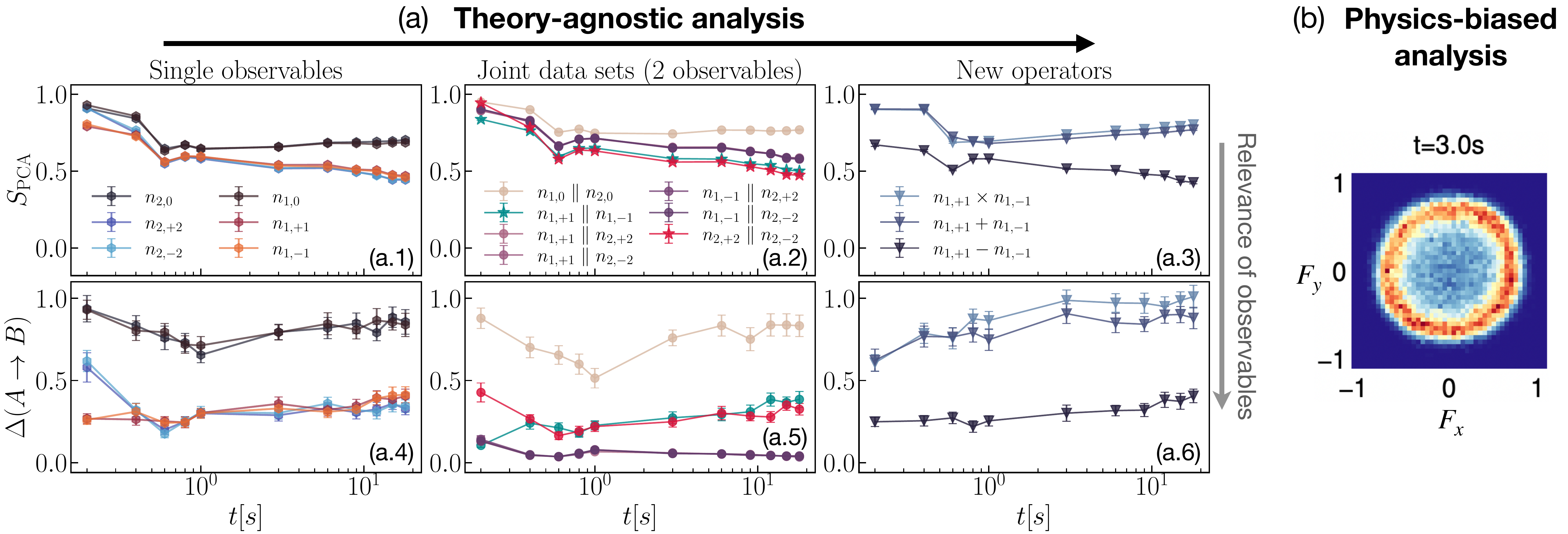}
	\caption{
	{ Assumption-free identification of relevant observables.}
        (a) PCA entropy, $S_\mathrm{PCA}$, and information imbalance, $\Delta(A \to B)$, as relevance metrics of physical observables:
        lower values of $S_\mathrm{PCA}$ signal  stronger correlations between the features of an observable, while lower values of $\Delta(A\to B)$ indicate that the features of a given observable (space $A$) are more informative of the full set of measured features (space $B$). 
        Both metrics clearly show that $n_{1, \pm 1}$ and $n_{2,\pm2}$ are more relevant (in the sense above) over the full evolution [(a.1) and (a.4)].
        Identification of relevant groups is also possible by analyzing joint data sets: for pairs of observables, $n_{1,+1} \mathbin\Vert n_{1,-1}$ and $n_{2,+2}$ $\mathbin\Vert$ $n_{2,-2}$, have the lowest $S_\mathrm{PCA}$ [$\star$ markers in (a.2)] (see ranking of all possible pairs in Appendix~\ref{app:relevance}). 
        Features from both `relevant' pairs are in fact required to better describe  space $B$ [points with $\Delta(A\to B)\approx 0$  in  (a.5), for which $A$ is defined  by the set of features of  $n_{1,\pm1} \mathbin\Vert n_{2,\pm2}$. Note that these points are almost on top of each other.] 
        Relevant new operators defined from the measured observables can also be identified, as illustrated here for a few combinations of $n_{1,+1}$ and $n_{1,-1}$, with $n_{1,+1}-n_{1,-1}$ being the most relevant [(a.3) and (a.6)]. (b)  Histogram of the transverse spin variable in the $F_x-F_y$ plane at $t=3s$, featuring a ring-like structure. Based on physical arguments~\cite{Prufer2018, Prufer2020, RevModPhys.85.1191, KAWAGUCHI2012253}, this variable is the relevant field to describe the quenched system.
        Our theory-agnostic approach identifies the relevant observables from which this variable is inferred  [see Eq.~\eqref{eq:4}], hence cross-validating the latter analysis.}
	\label{fig:2}
\end{figure*}

At the final parameters of the quench, which places the system into the regime of the easy-plane ferromagnetic phase~\cite{RevModPhys.85.1191, KAWAGUCHI2012253}, these define the transverse spin field $F_\perp=F_x+\mathrm{i}F_y$~\cite{Sadler2006}.
Here, the interplay between energy offsets and spin interactions favor a finite transverse magnetization. During the dynamics the transverse spin field approaches its ground state distribution, which manifests itself in the formation of a ring in the transverse spin histogram after approximately~$1$--$3\,\text{s}$, as shown in Fig.~\ref{fig:2}(b). Nevertheless, in this regime the system is still highly excited and transverse spin phase excitations evolve dynamically in a self-similar fashion \cite{Prufer2018}.
Such relaxation dynamics is quite rich but complex, making a controlled microscopic characterization extremely challenging. 
In fact, the interpretation above is motivated by heuristic arguments. The key point we are interested in here is to obtain such description based solely on experimental observations, analyzed in a blind-folded manner. That is, we wish to extract essential descriptive elements (most important operators and complexity of the dynamics) without relying on any assumption.
We note that the structure of the experimental setup described above is that of a continuous-variable quantum simulator (see, for instance, Refs.~\cite{Viermann2022, doi:10.1073/pnas.2301287120, Tajik2023}). In particular, the present experiment can be regarded as an analog quantum simulation of the out-of-equilibrium dynamics in a quantum field theory associated to the underlying physical system. Furthermore, since the spinor Bose gas under consideration features universal dynamics, this specific setup can be used to probe universal dynamics in a wide range of systems that share the same universal features.


Let us now describe the structure of the collected experimental data sets analysed in this work. 
At each evolution time, each density is sampled linearly at $p=184$ spatial locations along the longitudinal trap direction. 
Such measurements are repeated so as to gather $N_r=225$ independent realizations. 
We denote a single realization of a spatial density profile by a $p$-dimensional vector $\vec{n}^i_{\alpha}(t)$, where the considered internal state is succinctly labeled  by  ${\alpha}\equiv (F, m_F)$.
Thus, for each observable (density), and at each evolution time $t$, we obtain a data set ${\bf M}_{\alpha}(t)=\{\vec{n}^1_{\alpha}(t), \vec{n}^2_{\alpha}(t), \dots, \vec{n}^{N_r}_{\alpha}(t)\}$, which can be represented as a $(N_r \times p)$ rectangular data matrix. 
Using the terminology of the previous section, the features of the experimental data at hand are therefore the measured densities at selected positions. 
Examples of single realizations at different evolution times are shown in appendix~\ref{app:observables}.
Further, we also consider {\it joint} data sets formed by concatenating  horizontally data sets of the measured densities for different $\alpha$ at a given time. 
More specifically, each row in a joint data matrix is formed by appending, one after the other, single realizations of the observables of choice. 
Thus, for example, a joint data set of two observables will have twice as many features as the data set of one individual observable, but the same number of rows $N_r$. 
The particular order in which we concatenate the combined observables is irrelevant for our methods.
When needed, we will simply specify joint data sets by using the symbol of the corresponding observables joined by ``$\mathbin\Vert$''.

\section{Identification of relevant observables} \label{sec:relevance}

We now perform a descriptive analysis of the data sets above, with the goal of identifying  \emph{relevant} observables, that is, those observables that capture dominant spatial correlations across the evolution of the system.
This is, in fact, a crucial task in order to determine  good degrees of freedom emerging from the underlying physical system. 

The framework presented here builds on the complementary tools discussed in Secs.~\ref{subsec:S_PCA} and \ref{subsec:IB}.
We first compute the PCA entropy of the measured spatial density profiles and their combinations, and use it as a direct probe of the spatial correlations captured by those observables at each evolution time. 
Next, we use information imbalance as a way to determine which observables retain more information from the full space of observations, thereby providing a complementary metric for observable relevance. 
Let us note that, as discussed in Sec.~\ref{subsec:S_PCA}, $S_\mathrm{PCA}$ cannot reveal the presence of correlations if the embedding manifold of the data is curved. On the other hand, the estimator of the information imbalance in Eq.~\eqref{eq:2} depends only on the local neighborhood of each data point, and hence is well-suited to deal with arbitrarily non-linear manifolds~\cite{10.1093/pnasnexus/pgac039}.
In this sense, ranking the relevance of observables with both techniques provides a way to cross-verify the validity of our results. 
Our main results are shown in Fig.~\ref{fig:2}(a).
Let us first analyze the results for the PCA entropy [panels (a.1)--(a.3)].
A clear separation between two groups of observables is noted as the system evolves [panel (a.1)], with $n_{1,\pm1}$ and $n_{2,\pm2}$ having lower values of $S_\mathrm{PCA}$.
We conclude that these observables capture stronger spatial correlations and are hence more relevant in the sense specified above. 
Next, we consider joint data sets of two  observables [panel (a.2)].
The most relevant pairs according to this analysis are $\{n_{1,+1}, n_{1,-1}\}$  and $\{n_{2,+2}, n_{2,-2}\}$. 
The latter result is in excellent agreement with the physics-motivated analysis, in which such observables play a key role in the definition of the transverse spin [see Eq.~\eqref{eq:4} and Fig.~\ref{fig:2}(b)]. 
Going one step further,  in panel (a.3) we explore concrete functional combinations of the pair of observables $\{n_{1,+1}, n_{1,-1}\}$ (similar results are found for  $\{n_{2,+2}, n_{2,-2}\}$), which define new operators.
We find that $n_{1,+1}- n_{1,-1}$ has the lowest $S_\mathrm{PCA}$, once again in agreement with the physics-motivated {\it ansatz} [Eq.~\eqref{eq:4}]. 
We now turn our attention to the information imbalance analysis [panels (a.4)--(a.6)], which provides a complementary view on the relevance of observables.  
Here, $A$ refers to the space of features associated to a given observable, while $B$ is the full space of ($184\cdot 6=1104$) measured features.  
In panel (a.4), we can see that the observables with a lower PCA entropy have also a lower information imbalance. In other words, the observables that capture dominant spatial correlations are also more informative of the full space of data features (in the information imbalance sense). 
Interestingly, in the analysis of pairs of observables [panel (a.5)], we note that in order to describe the full space of observations, one needs to consider features from both of the relevant pairs. Indeed, we see that the space $A$ that combines features from the observables $n_{1,\pm1}$ and $n_{2,\pm2}$ have $\Delta(A \to B)\approx 0$, for the full evolution [note that the points corresponding to the four possible combinations of these observables are almost on top of each other in panel (a.5)].
The new operator $n_{1,+1}- n_{1,-1}$ has also a significantly lower information imbalance than the other considered operators in panel (a.6).

In Appendix~\ref{app:observables}, we show some instances of single realizations of the measured density profiles. While one can see a certain correlation in the behavior of the identified relevant observables (at the level of single realizations), our analysis rules out, in a systematic and unbiased way, the presence of other important correlations among the measured quantities. 
Further results for the analysis carried out here are shown in Appendix~\ref{app:relevance}. 
\section{Algorithmic complexity of the quantum dynamics}
\label{sec:complexity}

Finally, we  provide a further characterization of the data sets. 
Specifically, we study their intrinsic dimension $I_d$, at the considered evolution times. 
This allows us to characterize the \emph{algorithmic complexity}  of the quantum dynamics probed in our experiment. 
We note that this notion of complexity is fundamentally different from \emph{computational complexity} - prominent examples of the latter being entanglement~\cite{PhysRevLett.100.030504,PhysRevB.96.085107, PhysRevB.96.020408}  and quantum circuit complexity~\cite{ PRXQuantum.2.030316, Haferkamp2022}.
In the context of quantum many-body systems, computational complexity deals in general with characterizing the number of classical resources needed to efficiently simulate a quantum state (e.g., the bond dimension of a matrix product state representation or the number of gates needed in a quantum circuit to describe a target state). On a very distinct note, algorithmic complexity quantifies the notion of compression of information in a classical object, which in our particular case refers to the classical encoding of a quantum state given by the output of measurements (or even classical calculations).
We note that these two notions of complexity do not necessarily align with each other, and while the computational complexity has been explored intensively in the realm of strongly interacting systems, much less is known about the algorithmic complexity  of quantum many-body states.

As explored in recent studies on critical phenomena in many-body systems~\cite{PhysRevX.11.011040, PRXQuantum.2.030332, PhysRevB.106.144313, 2023arXiv230113216M}, the algorithmic complexity does provide a  physical picture of the complexity  of many-body states, revealing for example  an emergent simplification of the data manifold in systems featuring universality, where the physics also becomes parametrically simpler and can be described by a handful of universal exponents and functions. The present work thus extends the study of this kind of complexity to the realm of far-from-equilibrium quantum many-body dynamics.


\begin{figure}[bt!]
	\centering
	\includegraphics[width=1.\columnwidth]{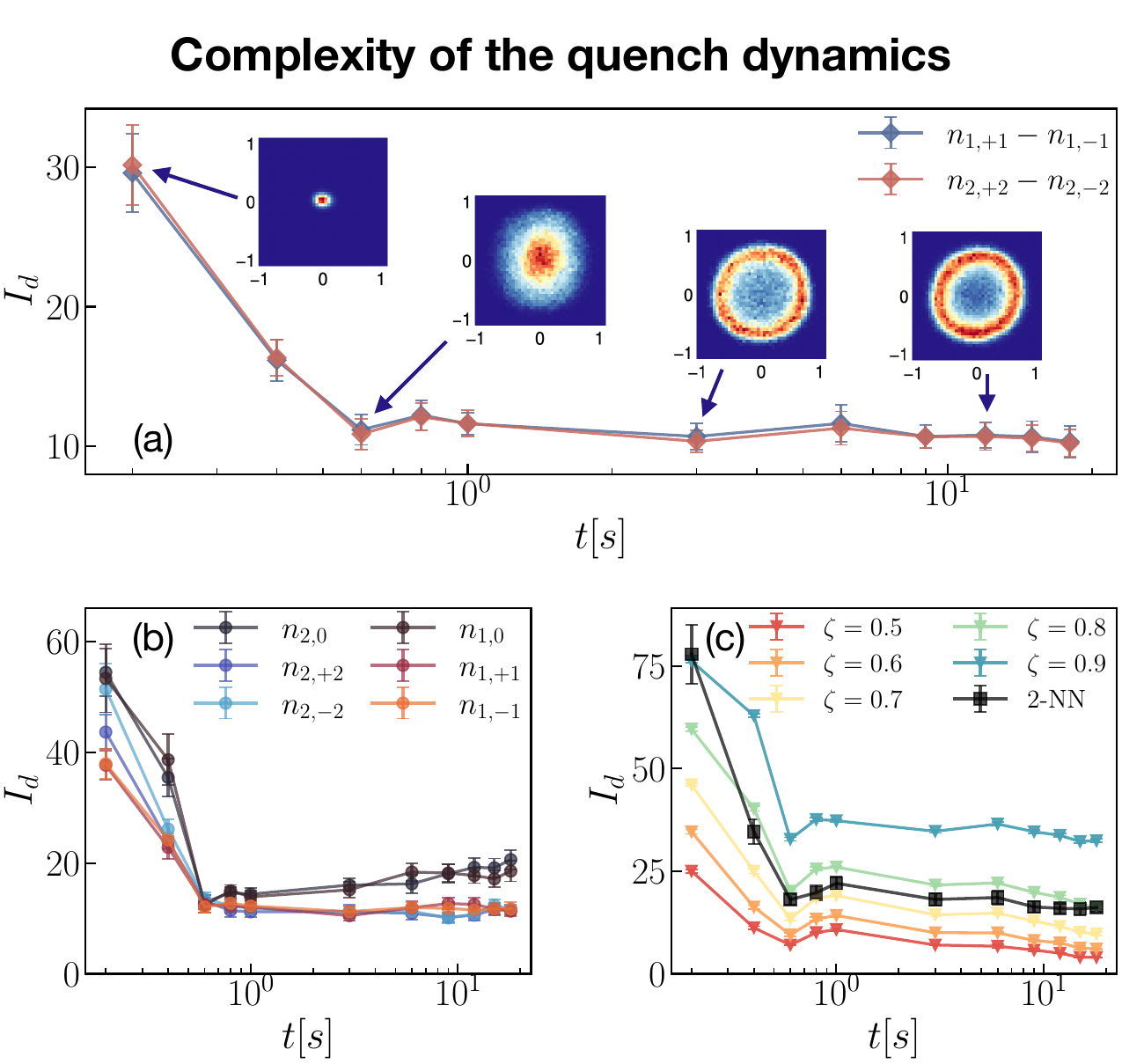}
	\caption{
        Intrinsic dimension as a function of time for (a) the relevant observables, $n_{1,+1}- n_{1,-1}$ and $n_{2,+2}-n_{2,-2}$, (b) all measured densities individually, and (c) joint data sets of all six observables together.
        In all instances, an initially large $I_d$ quickly decays to smaller values (around $t=0.6s$), subsequently exhibiting long plateaus. 
        The insets in panel (a) show histograms of the transverse spin in the $F_x-F_y$ plane at selected times. The first drop in the $I_d$ is associated to a grow in the spin length, which remains approximately constant for $t \gtrsim 1s$. In the latter regime a ring-like structure is then observed (illustrated here at $t=3s, 12s$). Spatial correlations of spin phase excitations exhibit self-similar dynamics in a regime that starts around $t\gtrsim  3s$~\cite{Prufer2020}. 
        The observed structural simplification of the data strongly correlates with such universal dynamics. 
        Hence, the plateaus in the plot of $I_d$ provide a lower bound for the onset of simpler dynamics and universal scaling.
        Panel (c) also shows the $I_d$ estimate based on PCA for various values of the cutoff $\zeta$ (see main text).
        }
	\label{fig:3}
\end{figure}

Shown in Fig.~\ref{fig:3}(a) is the plot of $I_d$ as a function of time, of the data sets corresponding to the identified relevant operators, 
$n_{1,+1}-n_{1,-1}$ and $n_{2,+2}-n_{2,-2}$. We observe the same trend in both instances: a quick decay of $I_d$ to considerably smaller values, subsequently displaying long, stable plateaus. 
The reduction of the $I_d$ signals a simplification of the data structure due to the buildup of correlations among the input variables. 
The latter is a direct manifestation of the correlations among the elementary constituents of the system. 
From the physical viewpoint, the post-quench correlations are associated with the formation of a ring-like structure, with approximately constant radius, in the transverse component of the collective spin degree of freedom (insets)~\cite{Prufer2018, Prufer2020}.
In turn, spatial correlations of the spin phase excitations exhibit universal scaling dynamics~\cite{Prufer2018}. 
In the present experiment, the universal scaling regime starts approximately at $t\sim 6s$~\cite{Prufer2020}. 
The physical basis for such scaling evolution is a dynamical reduction of the relevant parameters in the system. 
This is strongly consistent with the observed structural simplification of the data, as also observed in recent studies of critical behavior---in and out of equilibrium---in classical and quantum statistical mechanics systems~\cite{PhysRevX.11.011040, PRXQuantum.2.030332, PhysRevB.106.144313, 2023arXiv230113216M}. 
Therefore, in the present case, the observed $I_d$ plateau  provides a theory-agnostic lower bound for the timescale after which the dynamics may become simpler, allowing for the emergence of self-similar behavior.
Importantly, this prediction can be made by directly studying the $I_d$ of data sets of all measured densities, as shown in Fig.~\ref{fig:3}(b), where we observe an overall similar trend.
We note, however, that the ``irrelevant'' observables $n_{1,0}$ and $n_{2,0}$, have a growing $I_d$, rather than a plateau. 
This further confirms the relevance predictions based on PCA entropy and information imbalance.
Further, in Fig.~\ref{fig:3}(c) we plot the $I_d$ of joint data sets of the six measured observables together, showing once again the noted trend. 
In this plot, we also show an $I_d$ estimation based on PCA, which is defined by choosing an \emph{ad hoc} cutoff parameter $\zeta$, for the integrated spectrum of the covariance matrix \cite{doi:https://doi.org/10.1002/0470013192.bsa501, PhysRevX.11.011040}, i.e., $\sum_{k=1}^{I_d} \Tilde{\lambda}_k \approx \zeta$. 
We find that for all considered values of $\zeta$, we recover the same qualitative features as the TWO-NN $I_d$-estimate. A quantitative agreement can also be achieved for a suitable choice of $\zeta$ in the range $0.7\le \zeta \le 0.9$, at the different evolution times.
Since the TWO-NN estimator only depends on local neighborhoods and is therefore well-suited to deal with curved manifolds, as opposed to the PCA method, this agreement further confirms the applicability of PCA in our previous analysis and implies that curvature effects in the data manifold are negligible.

In Appendix~\ref{app:Pareto}, we show some examples of  the linear fits used to estimate $I_d$ as prescribed by the TWO-NN algorithm; see Eq.~\eqref{eq:3}. Statistical error on the results presented in this work were computed using a version of the delete-$d$ Jackknife standard error estimator via a stochastic subsampling algorithm without repetitions~\cite{politis, shao&tu}, as detailed in Appendix~\ref{app:subsampling}.

\section{Conclusions} \label{sec:conclusions}

We have introduced an assumption-free method to diagnose and rank relevant correlations in the dynamics of out-of-equilibrium quantum systems. The method exploits the full spectrum of principal components, as well as recently developed techniques based on information imbalance. We have successfully identified the most relevant operators describing the dynamics of Bose Einstein condensates, confirming previous heuristic approaches (and thus, validating the physical relevance based solely on experimental observations). Utilizing manifold characterization methods, we have also found stable plateaus of the intrinsic dimension of the data sets corresponding to different times, thus providing bounds on the time frame realizing universal quantum dynamics. Our approach is immediately extended to other classes of quantum simulators---including fermion gases and lattice spin models---providing a flexible, assumption-free framework to discover physical phenomena, as well as to validate their functioning. Our work complements recent theoretical approaches with similar goals regarding the identification of relevant observables~\cite{Miles2021, PhysRevLett.127.150504} and characterizing the complexity of quantum dynamics~\cite{2023arXiv230113216M,  PhysRevB.106.L041110, PhysRevE.108.044128}.

\begin{acknowledgments}

We are grateful to G. Bianconi, B. Lucini, M. Marsili, K. Najafi, S. Pedrielli, and P. Zoller, for discussions and feedback on this and related works. 
This work was partly supported by the MIUR Programme FARE (MEPH), by QUANTERA DYNAMITE PCI2022-132919, by the PNRR MUR project PE0000023-NQSTI, by the French
National Research Agency via QUBITAF (ANR-22-PETQ-0004, Plan France 2030), by the Deutsche For\-schungs\-gemeinschaft (DFG, German Research Foundation) through SFB 1225 ISOQUANT -- 273811115, and GA677/10-1, as well as under Germany's Excellence Strategy -- EXC-2181/1 -- 390900948 (the Heidelberg STRUCTURES Excellence Cluster).
\end{acknowledgments}

\appendix
\section{Experimentally measured observables} \label{app:observables}

\begin{figure*}[bt!]
	\centering
	\includegraphics[width=1.\textwidth]{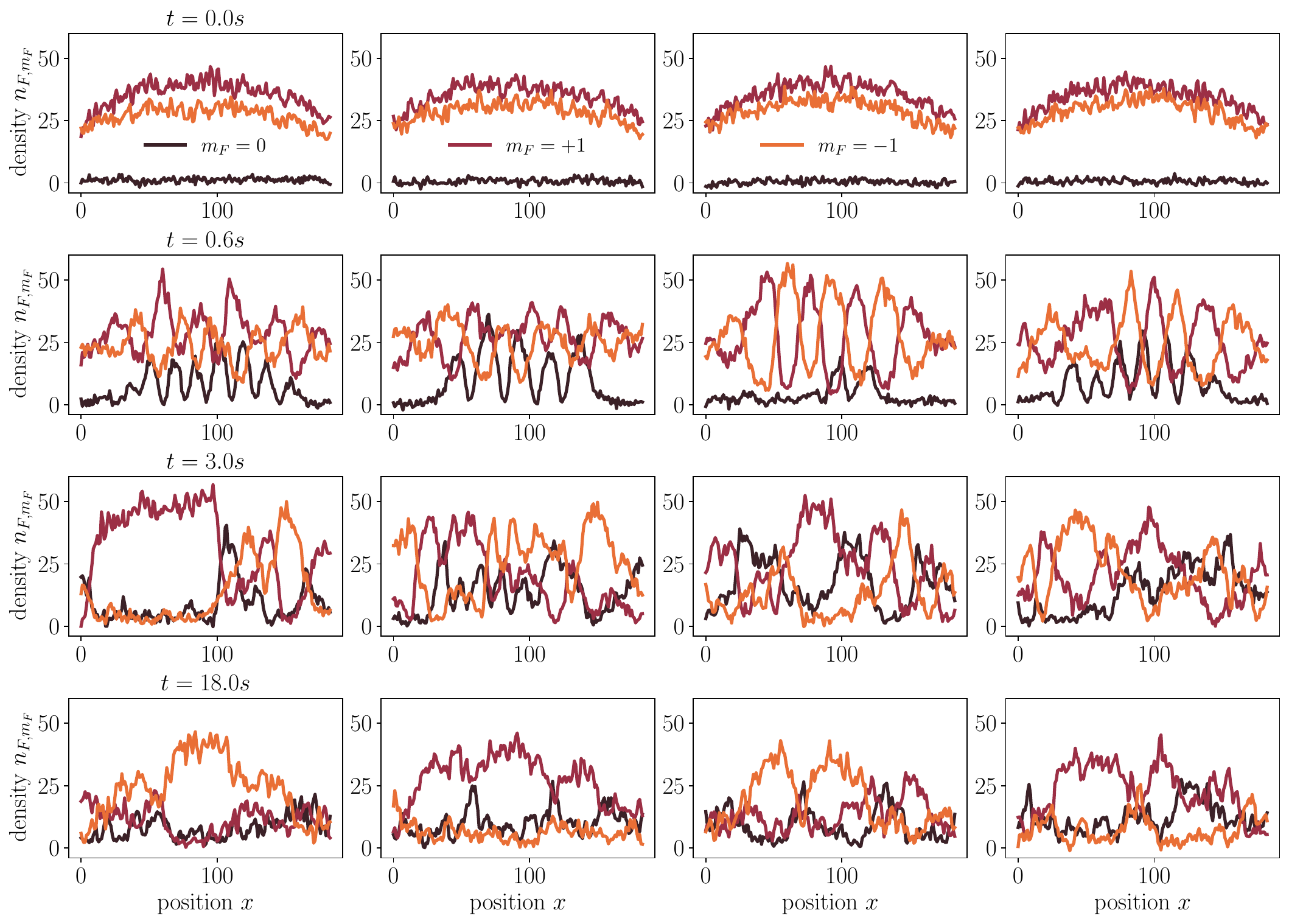}
	\caption{
	{ Experimentally observed density profiles in the hyperfine manifold $F=1$.}
        Single realizations of the measured densities are shown at selected evolution times in the subsequent rows.}
	\label{fig:appA_1}
\end{figure*}

\begin{figure*}[bt!]
	\centering
	\includegraphics[width=1.\textwidth]{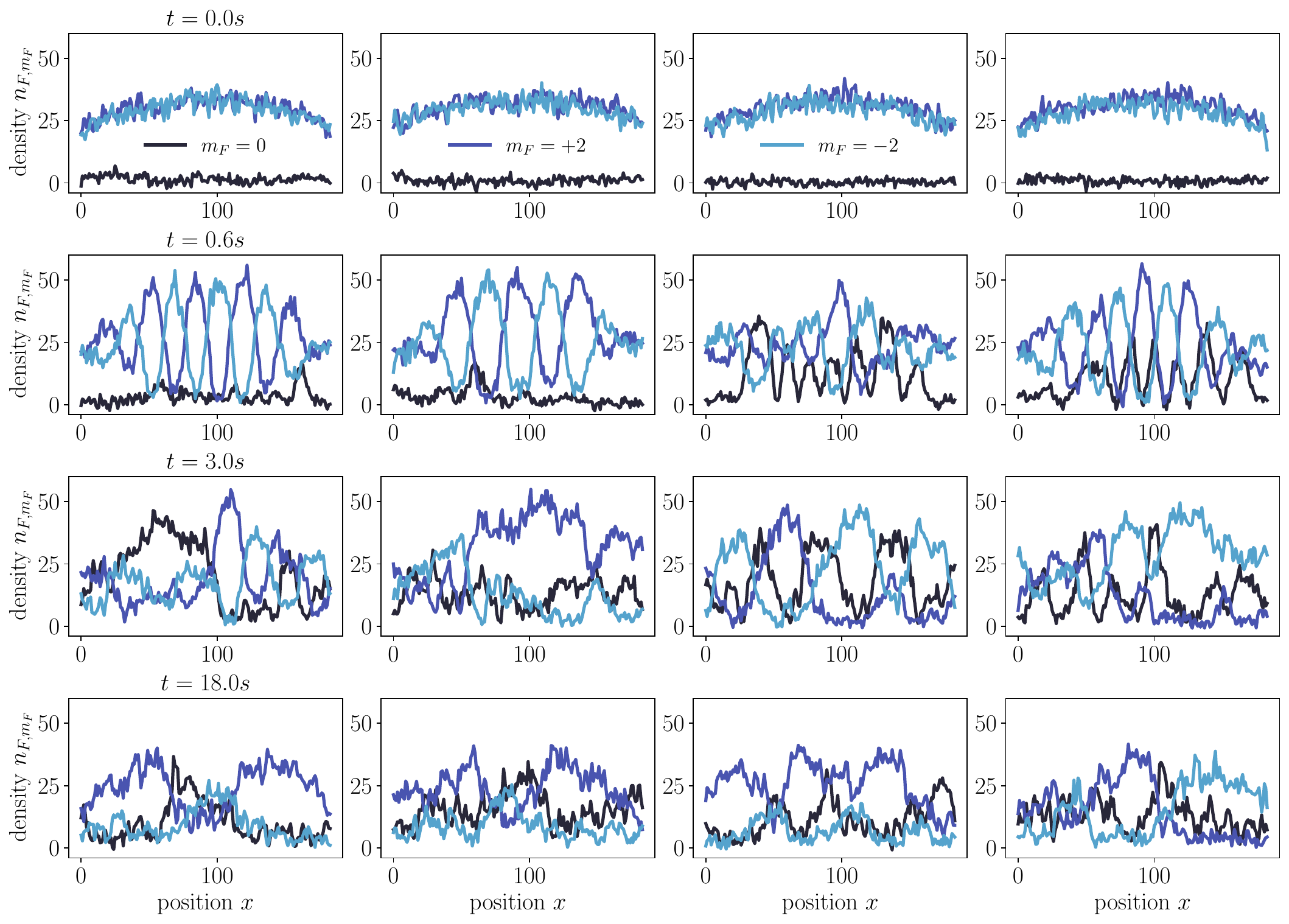}
	\caption{
	{ Experimentally observed density profiles in the hyperfine manifold $F=2$.}
        Single realizations of the measured densities are shown at selected evolution times in the subsequent rows.}
	\label{fig:appA_2}
\end{figure*}

In this appendix we show, for completeness, examples of single realizations of the measured density profiles at selected evolution times. We refer the reader to Ref.~\cite{Prufer2020}, where the experimental data analysed in this work were taken from, for further experimental details.
For the sake of clarity, we plot the density profiles corresponding to the hyperfine manifolds $F=1$ and $F=2$, separately in Figs.~\ref{fig:appA_1} and \ref{fig:appA_2}, respectively. 
The structure of the experimental data is as follows: before measuring the atomic densities a $\pi/2$ rotation around a transverse spin axis is performed. That means that measurements are done in the $x$- or $y$-basis. The states at the poles of the spin sphere, therefore, correspond to fully elongated spins along the $x$- or $y$-direction, respectively. At the poles one then measures all atoms in the internal states $(F,m_F)=(1,\pm 1)$ or $(F,m_F)=(2,\pm2)$~\cite{PhysRevLett.123.063603}.
In the initial state, all atoms are in the magnetic substate $m_F=0$, as discussed in the main text. However, due to the applied $\pi/2$ pulse before imaging the atoms, the atomic densities get modified and the initial $m_F=0$ population is split between the substates $m_F=\pm 1$ and $m_F=\pm 2$, for the readouts in $F=1$ and $F=2$, respectively.
This is precisely what we observe in the first row of in Figs.~\ref{fig:appA_1} and \ref{fig:appA_2}, which corresponds to $t=0.0s$.
We note however that no spatial correlations are imprinted in such an initial state. This is why both the PCA entropy and the intrinsic dimension of the measured densities are initially large (see Figs.~\ref{fig:2} and \ref{fig:3}, in the main text). That is, the corresponding data collections at short times exhibit the largest algorithmic complexity due to the statistical independence of the input variables (densities at different spatial positions).
At later evolution times, certain correlations between atomic densities in different substates can be observed at the level of single realizations. These are a consequence of certain symmetries of the system under study. For example, rotational invariance is ultimately responsible for a conserved population imbalance between the substates $m_F=\pm1$ during the spin-changing collisions that drive the spinor Bose gas out of equilibrium~\cite{RevModPhys.85.1191, KAWAGUCHI2012253}. However, excluding the presence of other important correlations by direct inspection becomes very hard and one needs to rely on techniques such as the ones introduced in this work, which allow to identify dominant correlations in a fully systematic and unbiased manner.


\section{Further results on relevant observables} 
\label{app:relevance}
\begin{figure*}[bt!]
        \label{fig:appB_1}
	\centering
	\includegraphics[width=1.\textwidth]{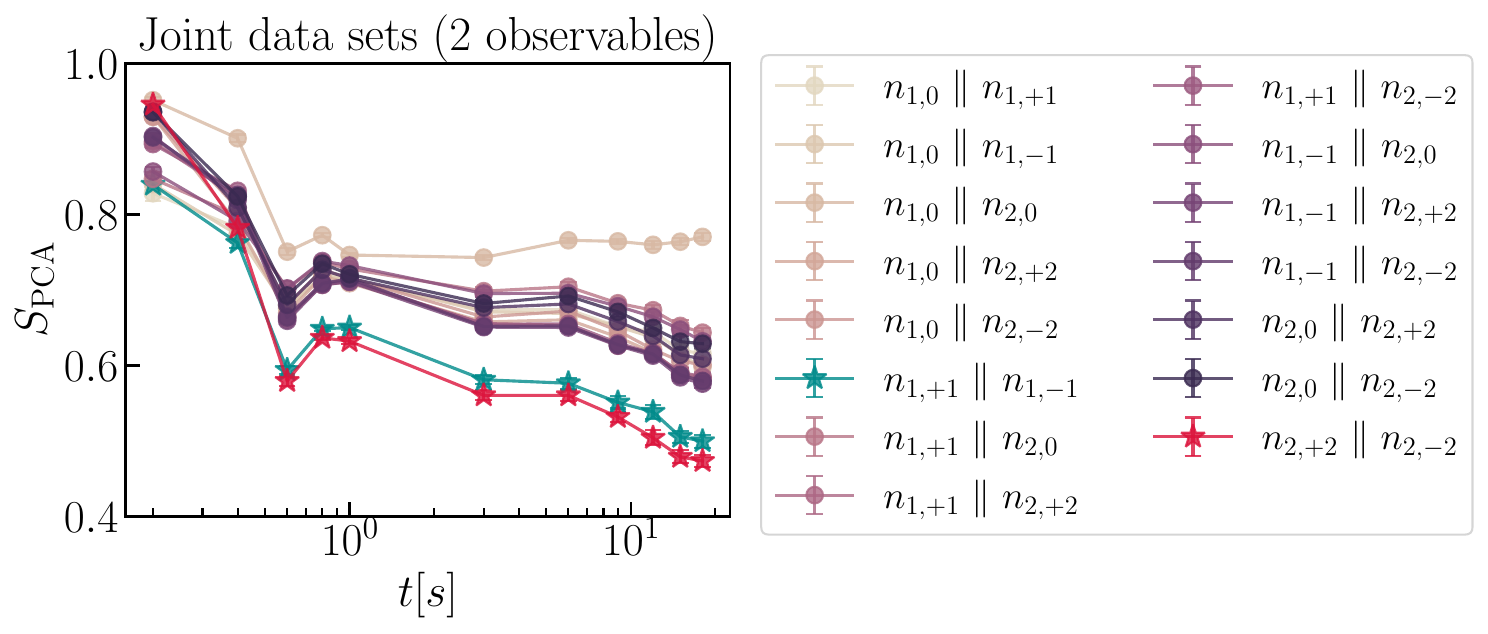}
	\caption{
	PCA entropy of joint data sets of two observables. This plot shows the results for all joint data sets of two observables [cf. Fig.~2(a.2) of the main text].
        }
\end{figure*}

\begin{figure*}[bt!]
        \label{fig:appB_2}
	\centering
	\includegraphics[width=1.\textwidth]{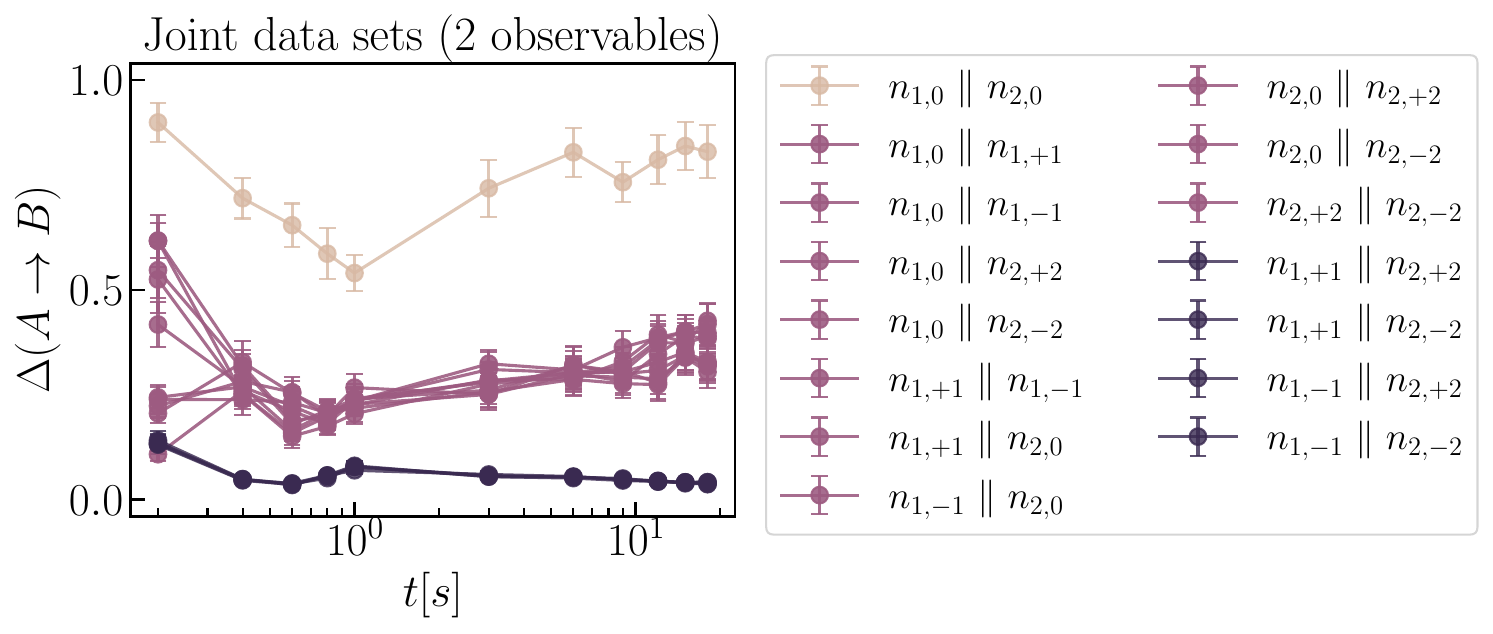}
	\caption{
	Information imbalance from the space of features of joint data sets combining two observables to the full space of measured features.  This plot shows the results for all joint data sets of two observables [cf. Fig.~2(a.5) of the main text]. Note that the points corresponding to the joint data sets  $n_{1,+1}\mathbin\Vert n_{2,+2}$, $n_{1,+1}\mathbin\Vert n_{2,-2}$, $n_{1,-1}\mathbin\Vert n_{2,+2}$, and $n_{1,-1}\mathbin\Vert n_{2,-2}$, lie basically on top of each other (they are equally informative), with an information imbalance $\Delta(A \to B)\approx 0$. These four data sets combine features of the two relevant pairs $\{ n_{1,+1}, n_{1,-1}\}$  and $\{ n_{2,+2}, n_{2,-2}\}$. 
        }
\end{figure*}

\begin{figure*}[bt!]
        \label{fig:appB_3}
	\centering
	\includegraphics[width=1.\textwidth]{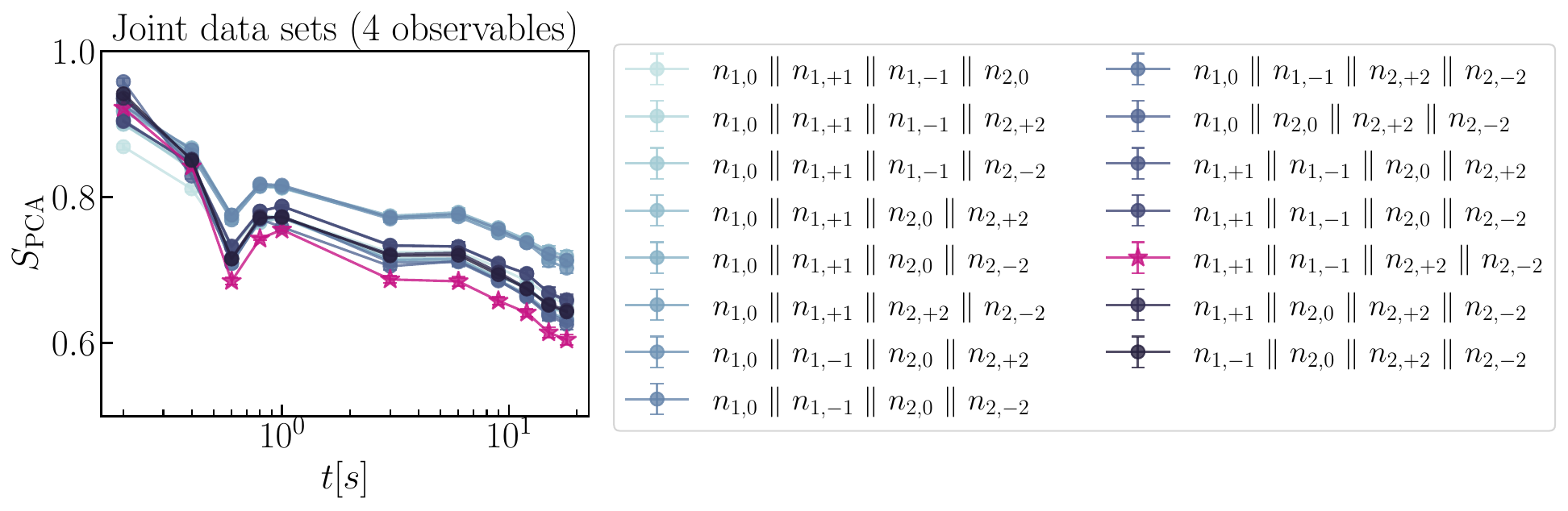}
	\caption{
	PCA entropy of joint data sets of four observables. The joint data set with the smallest PCA entropy is the one that combines the two relevant pairs ($\star$ markers).
        }
\end{figure*}

\begin{figure*}[bt!]
        \label{fig:appB_4}
	\centering
	\includegraphics[width=1.\textwidth]{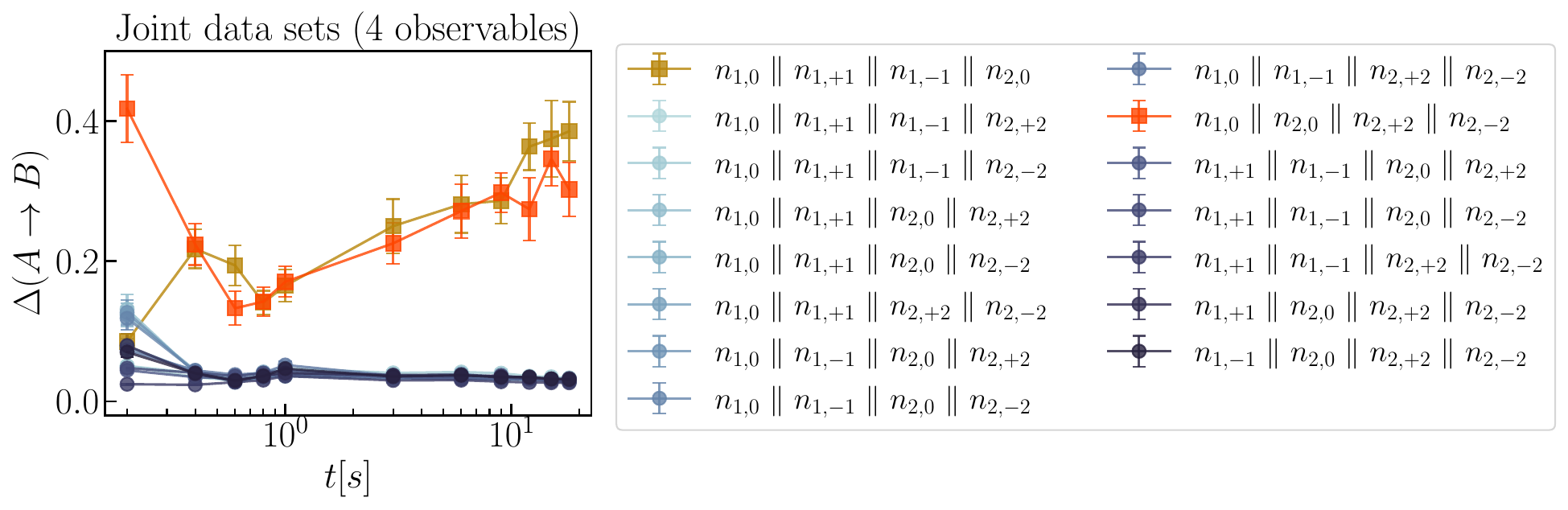}
	\caption{
	Information imbalance from the space of features of joint data sets combining four observables to the full space of measured features. All joint data sets that include features from the two relevant pairs $\{ n_{1,+1}, n_{1,-1}\}$  and $\{ n_{2,+2}, n_{2,-2}\}$ have a very small information imbalance. Instead, the joint data sets that only include features from one of the relevant pairs cannot predict so well the full space of features. The latter are indicated with square markers in this plot.  
        }
\end{figure*}

In this section, we show the ranking of operators beyond the ones discussed in the main text. In particular, in panels (a.2) and (a.5) of Fig.~\ref{fig:2}, we only show results for some of the possible combinations of two of the observed densities. 
In Figs.~~\ref{fig:appB_1} and \ref{fig:appB_2}, we show the results for all possible 15 combinations.
In these plots we can better appreciate the separation into three groups of combinations. Focusing first on $S_\mathrm{PCA}$, we can distinguish the two most informative pairs of observables with lower values of $S_\mathrm{PCA}$, throughout almost the whole evolution,  namely, $\{n_{1,+1}, n_{1,-1}\}$ and $\{n_{2,+2}, n_{2,-2}\}$ ($\star$ markers in Fig.~\ref{fig:appB_1}). These are follow by an set of combinations---each containing at least one of the observables in the identified relevant pairs---with intermediates values of $S_\mathrm{PCA}$, and finally the ``least'' informative combination $\{n_{1,0},n_{2,0}\}$.
The information imbalance in Fig.~\ref{fig:appB_2}, reveals a similar separation. First, with an almost zero information imbalance, we have the space of features corresponding to the combinations $\{n_{1,\pm 1}, n_{2,\pm2}\}$ (points with a darker color in Fig.~\ref{fig:appB_2}). This results, as discussed in the main text, means that in order to describe the full space of measured features we need to combine features from each of the two relevant pairs $\{n_{1,+1}, n_{1,-1}\}$ and $\{n_{2,+2}, n_{2,-2}\}$. The next group of combinations, yielding intermediate values of $\Delta(A \to B)$, contain features within only one of such relevant pairs, e.g.,  $\{n_{1,0}, n_{1,-1}\}$ or $\{n_{2,+2}, n_{2,-2}\}$. Finally, once again, we identify as the ``least'' informative combination that of  $\{n_{1,0},n_{2,0}\}$.
Similar observations are obtained  if one considers groups of more than two observables. This is illustrated here for groups of four observables (quadruplets) in Figs.~\ref{fig:appB_3} and \ref{fig:appB_4}. In terms of PCA entropy the most relevant combination is the one  that combines the features of the two relevant pairs, namely, the joint data set $n_{1,+1}\mathbin\Vert n_{1,-1} \mathbin\Vert n_{2,+2}\mathbin\Vert n_{2,-2}$ ($\star$ markers in Fig.~\ref{fig:appB_3}). We note however that in this case the relative difference in PCA entropy is not as pronounced as in the case of single or two observables. Regarding information imbalance, we observe once again those joint data sets that combine features from the two relevant pairs can predict almost entirely the full space of features. There are indeed only two combinations that only involve features from a single relevant pair (plus the two ``irrelevant'' observables $n_{1,0}$ and $n_{2,0}$), namely, $n_{1,0}\mathbin\Vert n_{1,+1} \mathbin\Vert n_{1,+1}\mathbin\Vert n_{2,0}$  and $n_{1,0}\mathbin\Vert n_{2,0} \mathbin\Vert n_{2,+2}\mathbin\Vert n_{2,-2}$ (square markers in Fig.~\ref{fig:appB_4}), which clearly have a significantly larger information imbalance compared to the rest. 

\section{Linear fit to estimate $I_d$ from the empirical cumulates in the TWO-NN method}\label{app:Pareto} 

\begin{figure*}[bt!]
	\centering
	\includegraphics[width=1.\textwidth]{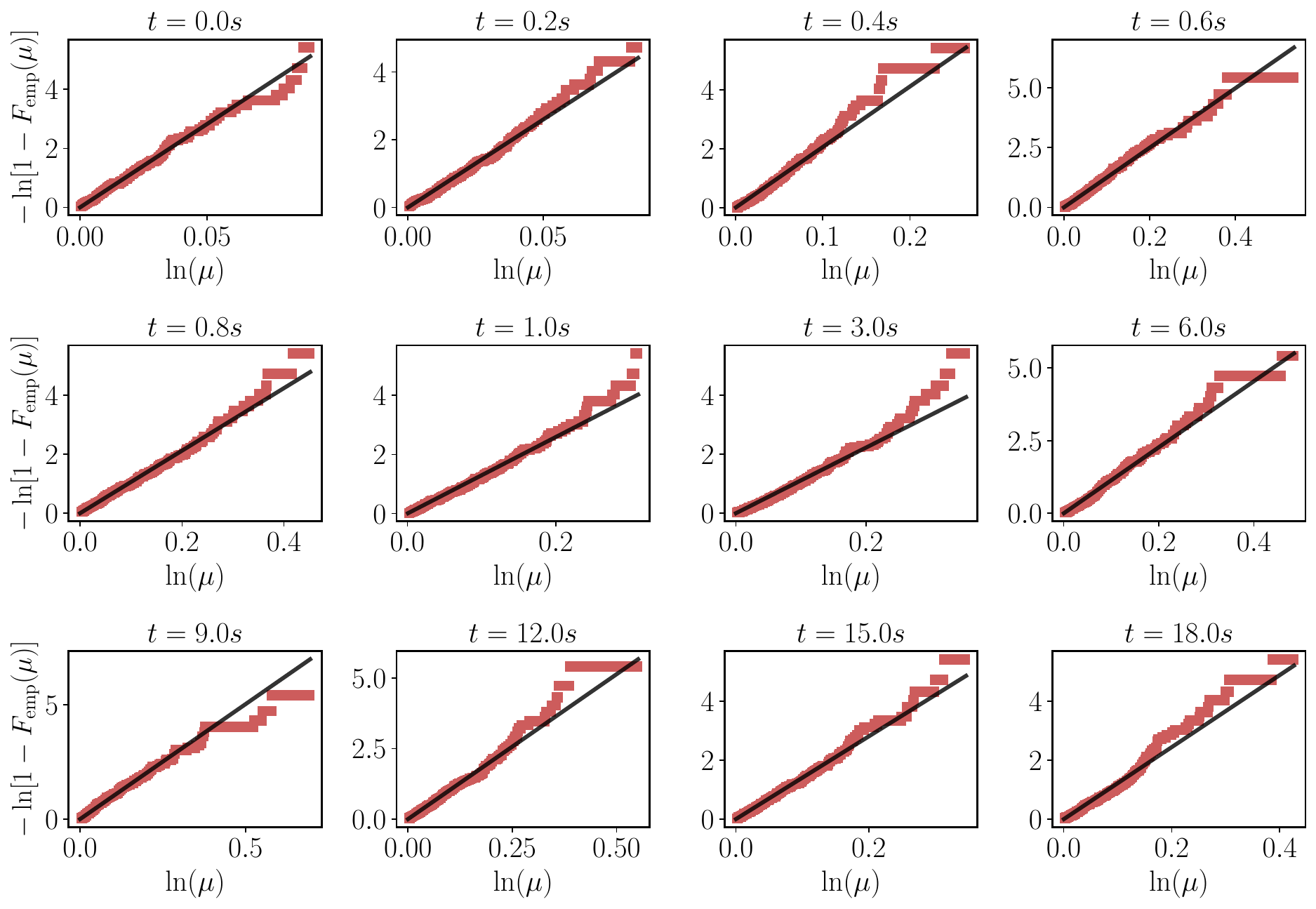}
	\caption{
	 Empirical cumulative distributions at all evolution times for the data sets corresponding to the relevant field $n_{2,+2}-n_{2,-2}$. The black curve show the linear fit  according to Eq.~\eqref{eq:3} in the main text, whose slope gives the estimated value of $I_d$. This procedure is valid as long as the empirical cumulative distribution function is consistent with a Pareto distribution, at least over a significant range of values of $\ln(\mu)$, as is clearly the case here.
        }
	\label{fig:appC}
\end{figure*}

In this section, we show examples of the linear fitting procedure used to estimate the value of the intrinsic dimension in the TWO-NN method; see Eq.~\eqref{eq:3} in the main text. In Fig.~\ref{fig:appC}, we show the empirical cumulative distributions of the ratios $\mu_i=r_{i_2}/r_{i_1}$, sorted in ascending order, for the observable $n_{2,+2} - n_{2,-2}$, at all evolution times. If the condition of constant density in the range of first two nearest neighbors holds, a plot of the resulting points $\{\ln(\mu), -\ln[1-F_\mathrm{emp}(\mu)] \}$ will be a line that passes through the origin and whose slope gives the estimated value of $I_d$. Verifying that the empirical cumulates are indeed consistent with a Pareto distribution as described above, is the first step to guarantee the applicability of the TWO-NN method. Besides, on its own, this kind of plot is also very informative about the local structure of complicated data manifolds.

\section{Subsampling error estimation} \label{app:subsampling}

Due to the limited number of  realizations used in the present analysis, we opted for using a  technique known as subsampling~\cite{politis, shao&tu} to have a sensible estimation of statistical errors. The subsampling algorithm is described below. 

At a given time and for a given measured observable, we have $N_r=225$ independent realizations forming our data set, that is, ${\bf M}= \{\vec{n}^i\}_{i=1}^{N_r}$, where for simplicity we have omitted the indices labeling the internal state and the evolution time. Using these data we compute a certain numerical statistic $\vartheta$. Given two preset integers $b$ and $q< N_r$, the subsampling analysis proceeds as follows: 

1. Form $b$ random `batches' (subsamples) of data by drawing $q<N_r$ points at random but \emph{without} replacement from  the data set ${\bf M}$.

2. Estimate the statistic of interest on each subsample, that is, $\vartheta_j$, for $j \in \{1, \dots, b\}$. 

3. Compute the mean of such  estimates $\overline{ \vartheta}  = \frac{1}{b}\sum_{j=1}^{b} \vartheta_j$. The standard error can  then be estimated as follows 
\begin{equation}
    \mathrm{SE} \approx \sqrt{\frac{q}{N_r-q}} \cdot \sqrt{\frac{1}{b}\sum_{j=1}^{b}(\vartheta_j - \overline{\vartheta})^2}.
\end{equation}
This formula is known as the delete-$d$ Jackknife standard error estimator (with stochastic subsampling)~\cite{politis, shao&tu}.

While formally this method requires $q/N_r\to 0$, $q\to \infty$, and $b\to \infty$ as $N_r\to \infty$ (so that the distribution of the $\vartheta_i$ converges to the sampling distribution of $\vartheta$), in practice the choice of these parameters is problem specific. In our analysis, we did not find significant changes for $b\ge 30$. Hence, we fixed $b=30$. Furthermore, to compute a meaningful statistic on each subsample, we set $q=100$, satisfying at least $q/N_r<1$. 

We used this method since sampling is performed \emph{without} replacement. This is important as the TWO-NN algorithm used to estimate the $I_d$ works under the assumption of no repetitions among the data points.

\bibliography{BIB}

\end{document}